\documentclass{article}
\usepackage{graphicx}
\usepackage{subfigure}
\usepackage{float}
\usepackage{amsfonts,amsbsy,amssymb,amsmath,amsthm,amsfonts,mathtools}
\usepackage{verbatim}
\usepackage{parskip}
\usepackage[T1]{fontenc}
\usepackage{multicol}
\usepackage{color}
\usepackage[autostyle]{csquotes}
\usepackage{pgf}
\usepackage{hyperref}
\usepackage{thmtools}
\usepackage{caption}
\usepackage{soul}
\usepackage{pdflscape}
\usepackage{relsize}
\usepackage{mathdots}
 
\declaretheoremstyle[%
  spaceabove=-6pt,%
 spacebelow=6pt,%
 headfont=\normalfont\itshape,%
 postheadspace=1em,
 qed=\qedsymbol%
]{mystyle} 

\hypersetup{
    colorlinks=true,
    linkcolor=blue,
    filecolor=blue,      
    urlcolor=blue,
    citecolor=blue,
    pdftitle={Overleaf Example},
    pdfpagemode=FullScreen,
    }

\newtheorem{thm}{Theorem}[section]
\newtheorem{prop}{Proposition}[section]
\newtheorem{cor}[thm]{Corollary}
\newtheorem{defn}[thm]{Definition}

\newtheorem{lemma}[thm]{Lemma}

\newtheorem{remark}[prop]{Remark}
\newtheorem{example}[thm]{Example}

\newcommand{\FF}{{\mathbb{F}}}

\newcommand{\CC}{\mathcal{C}}
\newcommand{\CD}{\mathcal{D}}

\newcommand{\ba}{\mathbf{a}}

\newcommand{\lcm}{\mathrm{lcm}}

\newcommand{\rk}{\mathrm{rank}}

\usepackage{authblk}

\begin{document}
\title{Trace duality and  additive complementary pairs of additive cyclic codes over finite chain rings}

\author[1]{Sanjit Bhowmick\thanks{This author would like to thank the Indian Institute of Technology Guwahati for its hospitality and support.}} 
\author[1]{Kuntal Deka}%
\author[2]{Alexandre Fotue Tabue\thanks{Supported by a "Research in Pairs" grant by CIMPA while his visit to the Institute of Mathematics, University of Valladolid, Spain.}} 
\author[3]{Edgar Mart\'inez-Moro\thanks{Partially supported by  y Grant  PID2022-138906NB-C21 funded by MICIU/AEI/10.13039/501100011033 and by ERDF/EU}}
\affil[1]{Department of Electronics and Electrical Engineering, Indian Institute of Technology Guwahati,
 Assam, 781039, India. sanjitbhowmick@rnd.iitg.ac.in; kuntaldeka@iitg.ac.in}
 \affil[2]{Department of Mathematics, University of Bertoua, Bertoua, Cameroon. alexfotue@gmail.com}
\affil[3]{Institute of Mathematics, University of Valladolid, Spain. 
edgar.martinez@uva.es}
 
\maketitle
\begin{abstract} 
This paper investigates the algebraic structure of additive complementary pairs of cyclic codes over a finite commutative ring. We demonstrate that for every additive complementary pair of additive cyclic codes, both constituent codes are free modules. Moreover, we present a necessary and sufficient condition for a pair of additive cyclic codes over a finite commutative ring to form an additive complementary pair. Finally, we construct a complementary pair of additive cyclic codes over a finite chain ring and show that one of the codes is permutation equivalent to the trace dual of the other.
\end{abstract}


\noindent\textbf{2020 AMS Classification Code:} 94B05, 15B05, 12E10.

\section{Introduction}\label{sec:intr}

Delsarte first studied additive codes over a finite field in 1971 \cite{MOO18}. Given a finite field $\mathbb{F}_{q^m}$ (where $q$ is a prime power) and a non-negative integer $n$, additive codes over $\mathbb{F}_{q^m}$ are subsets of $\mathbb{F}_{q^m}^n$ that are closed under addition but not necessarily under scalar multiplication by elements of $\mathbb{F}_{q^m}$. In particular, significant attention has been devoted to the class of additive codes that are closed under scalar multiplication by a subfield $\mathbb{F}_q \subset \mathbb{F}_{q^m}$, namely, $\mathbb{F}_q$-linear subspaces of $\mathbb{F}_{q^m}^n$. We will refer to them as  $\mathbb{F}_{q^m}|\mathbb{F}_q$ linear codes and the main theory on such codes can be found in \cite{Huff13}.

Now, if we consider a finite chain ring $R$, codes over $R$ are simply $R$-submodules of $R^n$. For a given subring $S \subset R$, we can also define additive codes in the same fashion as in the finite field case. If a code is closed under the cyclic shift, we refer to it as cyclic. There are a few works on cyclic additive codes over finite fields and finite chain rings; see, for example, \cite{SHI2022, VS24, MM18} and the references therein.

In the context of quantum error-correcting codes, this class of codes has attracted interest, especially when the alphabet is a quadratic extension $\mathbb{F}_{q^2}$ of $\mathbb{F}_q$. For instance, Ashikhmin and Knill \cite{AK21} constructed quantum codes using $\mathbb{F}_{q^2}|\mathbb{F}_q$ linear codes. Later, in \cite{KKKS06}, the authors revealed a deep connection between the existence of quantum error-correcting codes and $\mathbb{F}_{q^2}|\mathbb{F}_q$ linear codes endowed with a suitable inner product.

On the other hand, a pair of linear codes $\{\CC, \CD\}$ of length $n$ over a finite field $\mathbb{F}_q$ is called a \textit{linear complementary pair} (LCP) if $\CC \cap \CD = \{0\}$ and $\CC + \CD = \mathbb{F}_q^n$, that is, $\CC \oplus \CD = \mathbb{F}_q^n$. When $\CD = \CC^\perp$, the dual code of $\CC$, the code $\CC$ is referred to as a \textit{linear complementary dual} (LCD) code. LCD codes were first introduced by Massey in 1992 \cite{Mas92}, and the interest in both LCD and LCP codes has recently reemerged due to their applications in securing systems against side-channel and fault injection attacks \cite{BCC14,CG16}.
In this context, the security parameter when one uses an LCP $\{\CC, \CD\}$ is defined as $\min\{\mathrm{d}(\CC), \mathrm{d}(\CD^\perp)\}$, where $\mathrm{d}(\CC)$ denotes the minimum Hamming distance of the code $\CC$. In the LCD case, since $\CD^\perp = \CC$, the security parameter simplifies to $\mathrm{d}(\CC)$.

Carlet et al. \cite{CG18} proved that if $\{\CC, \CD\}$ is an LCP where both $\CC$ and $\CD$ are cyclic codes over $\mathbb{F}_q$, then $\CC$ is permutation equivalent to $\CD^\perp$. They further showed that this result extends to $2$D cyclic codes, provided the code length is relatively prime to the characteristic of $\mathbb{F}_q$ (i.e., the semi-simple case). Extending this result, Güneri et al. \cite{GOS18} showed that for abelian codes and the semisimple case, the equivalence $\CC \simeq \CD^\perp$ also holds.
This result is more general and can be viewed in the context of group codes; thus, the same result has been proven for an LCP of group codes without requiring any assumption on the characteristic of the field (i.e., without the need for semi-simplicity), see \cite{BORELLO2020}. Finally, this result also holds for LCP codes over finite chain rings \cite{Selcen2020}. Recently, a similar result was proven for an LCP of algebraic geometry codes in \cite{BDM23}.
 There are a few works on additive complementary dual codes; see, for example, \cite{choi23} and the references therein.

 In the present work, Linear Complementary Pairs (LCP) of additive cyclic codes over a finite commutative ring are investigated. The paper outline is as follows. Section~\ref{sec:pre} provides some preliminaries on additive codes and the trace duality. In Section~\ref{sec:structure}, we study the structure and polynomial definition of additive cyclic codes over a   Galois extension $S|R$ of finite chain rings. Section~\ref{sec2} deals with the description of the trace dual of the additive codes defined in Section~\ref{sec:structure}. Additive Complementary Pairs of codes (not necessarily cyclic ones) are studied in Section~\ref{sec:acp}, while in  Section~\ref{sec:acpc}, Additive Complementary Pairs of cyclic codes are tackled, providing a generalization of the result in \cite{CG18} for this type of codes.

\section{Preliminaries}\label{sec:pre}

A chain ring is a ring whose ideal lattice forms a chain. In this paper, $S$ and $R$ will denote finite commutative chain rings. We will denote  the  maximal ideal of  $S$ as $\mathfrak{m}_{S}$ and its nilpotency index as $e$. We say that $S$ is a ring extension of $R$, denoted $S|R$, if $R$ is a subring of $S$, $\mathfrak{m}_{R} = \mathfrak{m}_{S}\cap \mathcal{R}$, and $1_{R} = 1_{S}$. The extension $S|R$ is a Galois extension of degree $2$ if $S$ is isomorphic to the quotient ring $R[x]/\langle f(x) \rangle$, where $f(x)$ is a basic irreducible polynomial of degree $2$ over $R$. The Galois group $\texttt{Aut}_{R}(S)$ of this extension consists of those ring automorphisms of $S$ such that restricted to $R$ are the identity map of $R$.

We will denote by $R/\mathfrak{m} = \mathbb{F}_q$ and $S/\mathfrak{m} = \mathbb{F}_{q^2}$ the quotient of $R$ (respectively $S$) by its maximal ideal.  According to \cite[Theorem XV.2]{McDonal1974}, we have $\texttt{Aut}_{\mathbb{F}_q}(\mathbb{F}_{q^2}) \simeq \texttt{Aut}_{R}(S)$ and  $\rk_{R}(S) =[\mathbb{F}_{q^2}:\mathbb{F}_q]= |\texttt{Aut}_{R}(S)|$. Thus, the ring $S$ can be regarded as a free $R$-module of rank $2$.

For the chain ring $S$, the \emph{ Teichmüller set } $\mathcal T\subset S$ is the unique set of $q^2$ elements in $S$ such that the image of $\mathcal T$ under the canonical projection $S\mapsto S/\mathfrak{m} = \mathbb{F}_{q^2}$ is the entire field,
each element $t\in\mathcal T$ satisfies $t^{q^2}=t$, and $\mathcal T$ contains the multiplicative representatives of $\mathbb{F}_{q^2}$ in $S$.
If  $S|R$ is a Galois extension of finite chain rings of degree $2$, 
  then there exists an element $\xi \in S$ whose multiplicative order is $q^2 - 1$  \cite[Theorem 14.27]{ZWan11}. Then, given the following set is called the  \textit{Teichmüller set} of $S$
\[
\mathcal{T}_{S} := \{0, 1, \xi, \xi^2, \ldots, \xi^{q^2 - 2}.\}
\]
 One can define $\zeta = \xi^{\frac{q^2 - 1}{q - 1}}$, and hence $\zeta \in R$ has multiplicative order $q - 1$. Thus, it is easy to check that the set $ \mathcal{T}_{R} = \{0, 1, \zeta, \zeta^2, \ldots, \zeta^{q - 2}\}. $
 is the \textit{Teichmüller set} of $R$.

Let $\mathrm{Aut}_{R}(S)$ denote the group of all $R$-automorphisms of $S$. Since $S$ is a Galois extension of $R$ of degree $2$, it follows, by an argument similar to that in \cite[Corollary 14.33]{ZWan11}, that $\mathrm{Aut}_{R}(S)$ is a cyclic group of order $2$.

In fact, the map $\phi : S \rightarrow S$ defined by  
\begin{equation}
    \phi(\alpha) = \alpha_0 + \alpha_1 \xi^q,
\end{equation}
for $\alpha = \alpha_0 + \alpha_1 \xi$ with $(\alpha_0, \alpha_1) \in R^2$, is an automorphism of $S$ that fixes $R$. Moreover, $R$ is the largest subring of $S$ fixed pointwise by $\phi$. This automorphism $\phi$ is called the \textit{Frobenius automorphism} of $S$ over $R$, and it generates the cyclic group $\mathrm{Aut}_{R}(S)$.

 Let $\mathrm{Tr} : S\rightarrow R$ the map given by $s\mapsto s+\phi(s)$, for all $s\in S$. $\mathrm{Tr}$ is called \textit{the generalized trace} of $S$ relative to $R$. It is well known that $\mathrm{Tr}$ is a surjective $R$-module homomorphism. 
The following lemma will be useful later.

\begin{lemma}\label{lem:trace}
 The kernel of $\mathrm{Tr}$ is equal to $\mu R$, for some $\mu\in S\setminus R$ such that $\mathrm{Tr}(\mu)=0$, i.e., 
 \begin{equation}
 \mathrm{Ker}(\mathrm{Tr})=\mu R=\{\mu r\mid r\in R\}.    
 \end{equation}
\end{lemma}

\begin{proof}
 It is well known that $\mathrm{Tr}$ is a surjective $R$-module homomorphism. Therefore, by the first isomorphism module theory, we have $S/\mathrm{Ker}(\mathrm{Tr})\simeq R$. Consequently, it follows that $|\mathrm{Ker}(\mathrm{Tr})|=|R|$. Furthermore, we observe that for any $r\in R$, we get $\mathrm{Tr}(r)=r+\phi(r)=2r$, as $\phi$ is the Frobenius automorphism of $S$ over $R$. Thus, $\mathrm{Tr}(r)\neq 0$ for all $r\in R\setminus\{0\}$ as characteristic of $R$ is an odd prime that there is an element $\mu\in S\setminus R$ such that $\mathrm{Tr}(\mu)=0$. Let $x\in\mu R$, then $x=\mu r$, some $r\in R$. It follows  that $\mathrm{Tr}(x)=0$, for all $x\in \mu R$, hence, $\mu R=\mathrm{ker}(\mathrm{Tr})$. 
\end{proof}
Furthermore, for an element $\mu\in S\setminus R$ such that $\mathrm{Tr}(\mu)=0$, the set $\{1, \mu\}$ is a basis of $S$ over $R$, as $S$ is a free $R$-module.  

\begin{defn}
    A linear  code of length $n$ over $ R$ is just an $R$-submodule of $ R^n$. An additive code of length $n$ over $S|R$ is just an $R$-submodule of $S^n$.
\end{defn}

\begin{defn}
  For an additive code of length $n$ over $S|R$, the trace dual of $\CC$ is given by  
  \begin{equation}
      \CC^{\perp_{\mathrm{Tr}}}=\{\mathbf{a}\in S^n\mid\mathrm{Tr}(\mathbf{a}\cdot\mathbf c)=0~\text{for~all}~ \mathbf c\in \CC\}.
  \end{equation}
\end{defn}
Note that $\CC=\left(\CC^{\perp_{\mathrm{Tr}}}\right)^{\perp_{\mathrm{Tr}}}$ and $|\CC||\CC^{\perp_{\mathrm{Tr}}}|=|\mathcal{S}_n|$ (see \cite{Woo99}).
 
\section{Structure of additive cyclic codes}\label{sec:structure}
Throughout the paper, we will assume that the length of the codes $n$ will be a positive integer that is not divisible by the characteristic of the residue field $R/\mathfrak m=\mathbb F_q$; thus,  the polynomial $x^n-1$ is square-free in $\mathbb{F}_q[x]$ and $x^n-1$ admits a unique factorization into a product of pairwise coprime, basic irreducible polynomials in $R$ (resp. $S$).   We will denote the following polynomial quotient rings as  
 \begin{equation}\label{eq:rings}
     \mathcal{R}_n=R[x]/\langle x^n-1\rangle, \qquad
 \mathcal{S}_n=S[x]/\langle x^n-1\rangle .
 \end{equation}
 Both rings $\mathcal{R}_n$ and $\mathcal{S}_n$ are principal, see  \cite{NS2000}. 
 If we set $\overline{x}=x+\langle x^n-1\rangle$, the map
$$
\begin{array}{cccc}
  \Psi : & S^n & \rightarrow & \mathcal{S}_n \\
    & (a_0,\ldots, a_{n-1}) & \mapsto & \sum\limits_{j=0}^{n-1}a_j\overline{x}^j 
\end{array}
$$
is an $R$-module isomorphism. Moreover $\mathcal{R}_n$ is a free $R$-module of rank $n$, and $\mathcal{S}_n$ is a free $R$-module of rank $2n$.

\begin{defn}
 An additive cyclic code over $S|R$ of length $n$ is an $R$-submodule $\mathcal{C}$ of $ S^n$ that satisfies
 $$(c_0, c_1,\ldots, c_{n-1})\in \CC\implies (c_{n-1}, c_0, \ldots, c_{n-2})\in \CC.$$
\end{defn}

 It is easy to check that a (linear) cyclic code of length $n$ over $S$ can be seen as an $R$-submodule of $\mathcal{R}_n$, and an additive cyclic code of length $n$ can be represented as  an $R$-submodule of $\mathcal{S}_n$. We will follow this polynomial notation of (additive) cyclic codes in the rest of the paper. For more details on additive cyclic codes over chain rings, we refer to \cite{MM18} and the references therein.

\begin{defn}
  A nonempty subset $\mathcal{C}$ of $S^n$ is an $S|R$ additive cyclic code of length $n$  if and only if $\Psi(\mathcal{C})$ is an $\mathcal{R}_n$-submodule of $\mathcal{S}_n$.  
\end{defn}

In the sequel, we identify any free additive code of length $n$ over $S|R$ with an $\mathcal{R}_n$-submodule of $\mathcal{S}_n$.  
The quotient ring $\mathcal{S}_n$ is a free $\mathcal{R}_n$-module of rank two. Therefore, any $\mathcal{R}_n$-submodule of $\mathcal{S}_n$ is generated by at most two elements of $\mathcal{S}_n$. From now on,  recall also  that $\mathrm{Ker}(\mathrm{Tr})=\mu R$  (see \ref{lem:trace}).

\begin{lemma}\label{lem_repre} Let $\mathcal{C}$ be an  free $S|R$ additive code of length $n$. Then there exist unique monic divisors $f(x)$ and $g(x)$ of $x^n - 1$ in $R[x]$, and a polynomial $r(x) \in R[x]$, such that
\[
\mathcal{C} = \left\langle f(\overline{x}) + \mu \cdot r(\overline{x}) \right\rangle_{\mathcal{R}_n} \oplus \left\langle \mu g(\overline{x}) \right\rangle_{\mathcal{R}_n},
\]
where $f(x)$ divides $r(x)$ and $\deg(r(x)) < \deg(g(x))$.
\end{lemma}

\begin{proof} To prove the result, we first define a map
$$\begin{array}{cccc}
  \psi: & \CC & \rightarrow & \mathcal{R}_n \\
    & a(\overline{x})+\mu b(\overline{x}) & \mapsto & a(\overline{x}).
\end{array}$$   
 Note that $\psi$ is an $\mathcal{R}_n$-module homomorphism. Therefore, $\psi(\mathcal{C})$ is an ideal of $\mathcal{R}_n$.
 Since $\mathcal{C}$ is free (as $R$-module), there is a free submodule $\mathcal{C}_1$ of $\mathcal{C}$ such that the restriction of $\psi$ to $\mathcal{C}_1$ is an $R$-module isomorphism. Thus $\psi(\mathcal{C})$ is a free cyclic code over $R$. Then there exists a unique divisor $f(x)$ of $x^n-1$ in $R[x]$  such that  $\psi(\mathcal{C})=\langle f(\overline{x})\rangle_{_{\mathcal{R}_n}}$.  
Consequently, $\mathcal{C}_1$ is also cyclic. Thereby $\mathcal{C}_1 = \left\langle f(\overline{x}) + \mu r(\overline{x}) \right\rangle_{\mathcal{R}_n} $ where $r(x)\in R[x]$.
Besides, $\mathcal{C}=\mathcal{C}_1\oplus\mathrm{Ker}(\psi)$ and $\mathrm{Ker}(\psi)$ is free as $R$-module. Now,
$$
\mathrm{Ker}(\psi)=\{ a(\overline{x})+\mu b(\overline{x})\in \CC\mid a(\overline{x})=0\}.
$$
If we define the set $A=\{b(\overline{x})\in \mathcal{R}_n \mid \mu b(\overline{x})\in \mathrm{Ker}(\psi)\}$, then $\mathrm{Ker}(\psi)=\mu A$. We have $A$ is an ideal of the principal ideal ring $\mathcal{R}_n$. Therefore, there exists a unique monic divisor $g(x)$ of $x^n-1$ in $R[x]$ such that $A=\langle g(\overline{x})\rangle_{_{\mathcal{R}_n}}$.  Thus, $$\CC=\langle f(\overline{x})+\mu r(\overline{x}), \mu g(\overline{x}) \rangle_{_{\mathcal{R}_n}}$$ for some polynomial $r(x)$ in $R[x]$. 
If $\deg(r(x))\geq \deg(g(x))$. Then, by division algorithm, there exist $s(x)$ and $u(x)$ in $R[x]$ such that $r(x) =u(x)g(x) +s(x) $ with $\deg(s(x))<\deg(g(x))$. Thus, 
 \[
\CC=\langle f(\overline{x})+\mu r(\overline{x}), \mu g(\overline{x}) \rangle_{_{\mathcal{R}_n}}=\langle f(\overline{x})+\mu s(\overline{x}), \mu g(\overline{x}) \rangle_{_{\mathcal{R}_n}}.  
 \]
 Hence, we may consider $\deg(r(x))< \deg(g(x))$. Finally, $h(\overline{x})(f(\overline{x})+\mu r(\overline{x}))=\mu h(\overline{x})r(\overline{x})\in \mathcal{C}_1\cap \mathrm{Ker}(\psi)=\{0\}$, where $f(x)h(x)=x^n-1$. It follows that $x^n-1$ divides $h(x)r(x)$. Thus, $f(x)$ divides $r(x)$, since $f(x)$ and $h(x)$ are coprime.
\end{proof}

\begin{thm}\label{th-3.3} Let $\mathcal{C}$ be a free $S|R$ additive code of length $n$. Then there exist unique monic divisors $f(x)$ and $g(x)$ of $x^n-1$ in $R[x]$  and a polynomial $r(x)$ in $R[x]$ with $\deg(f(x)r(x))<\deg(g(x))$ for which 
$$\mathcal{C} = \langle f(\overline{x})(1 +\mu r(\overline{x}))\rangle_{_{\mathcal{R}_n}}\oplus\langle \mu g(\overline{x}) \rangle_{_{\mathcal{R}_n}}.$$ Moreover $S_1\cup S_2$ is an $R$-basis of $\mathcal{C}$, where \begin{eqnarray*}
S_1&:=&\{\overline{x}^if(\overline{x})(1+\mu r(\overline{x}))\mid 0\leq i < n-\deg(f(x))\};\\
S_2&:=&\{\overline{x}^j\mu g(\overline{x})~: ~ 0\leq j < n-\deg(g(x))\}.  
\end{eqnarray*}
\end{thm}
 
\begin{proof} Let $\mathcal{C}$ be a free additive code of length $n$ over $S|R$. By Lemma \ref{lem_repre}, there exist unique monic divisors $f(x)$ and $g(x)$ of $x^n - 1$ in $R[x]$, and a polynomial $r_0 (x) \in R[x]$, such that
\[
\mathcal{C} = \left\langle f(\overline{x}) + \mu r_0(\overline{x}) \right\rangle_{\mathcal{R}_n} \oplus \left\langle \mu g(\overline{x}) \right\rangle_{\mathcal{R}_n},
\]
where $f(x)$ divides $r_0(x)$ and $\deg(r_0(x)) < \deg(g(x))$. Besides, the sets $$\{\overline{x}^if(\overline{x})(1+\mu r_0(\overline{x}))\mid 0\leq i < n-\deg(f)\}$$ and $\{\overline{x}^j\mu g(\overline{x})~: ~ 0\leq j < n-\deg(g)\}$ are the $R$-bases of $\left\langle f(\overline{x}) + \mu r_0(\overline{x}) \right\rangle_{\mathcal{R}_n}$ and $\left\langle \mu g(\overline{x}) \right\rangle_{\mathcal{R}_n}$, respectively. Therefore, $S_1\cup S_2$ is an $R$-basis of $\mathcal{C}.$
Since $f$ divides $r_0$ implies $r_0=fr$ where $r\in R[x]$ and $\deg(r)<\deg(g)-\deg(f)$.
\end{proof}

\section{Trace duality}\label{sec2}

 Recall that $\mathcal{S}_n$ is a free $R$-module of rank $2n$. The $\star$-inner and $\circledast$-inner product on $\mathcal{S}_n$, defined by
\begin{equation}
 \left(\sum_{i=0}^{n-1} a_i \overline{x}^i\right)\star \left(\sum_{i=0}^{n-1} b_i \overline{x}^i\right) = \sum_{i=0}^{n-1} a_i b_i,   
\end{equation}
and
\begin{equation}
  \left(\sum_{i=0}^{n-1} a_i \overline{x}^i\right) \circledast \left(\sum_{i=0}^{n-1} b_i \overline{x}^i\right) = \operatorname{Tr}\left(\sum_{i=0}^{n-1} a_i b_i\right),  
\end{equation}
 
are non-degenerate symmetric bilinear forms over $S$ and $R$, respectively. The binary operation $\circledast$ on $\mathcal{S}_n$ is called \textit{the trace} over $\mathcal{S}_n$. Note that $$\alpha \textbf{u}(\overline{x})\circledast \beta\textbf{v}(\overline{x})=\mathrm{Tr}(\alpha\beta)(\textbf{u}(\overline{x})\star\textbf{v}(\overline{x})),$$ for all $(\textbf{u}(\overline{x}), \textbf{v}(\overline{x}))$ in $(\mathcal{R}_n)^2$ and $(\alpha, \beta)\in  S^2$.

\begin{remark} Let $(a(\overline{x}),  b(\overline{x}), a'(\overline{x}),  b'(\overline{x}))\in(\mathcal{R}_n)^4.$ Then 
$$(a(\overline{x})+\mu b(\overline{x}))\circledast(a'(\overline{x})+\mu b'(\overline{x}))=2(a(\overline{x})\star a'(\overline{x})+\mu^2(b(\overline{x})\star b'(\overline{x}))).$$
\end{remark}

\begin{defn}  Let $\mathcal{C}$ be an $S|R$ additive code of length $n$. The trace dual of $\mathcal{C}$, denoted $\mathcal{C}^{^{\perp_{_\mathrm{Tr}}}}$, is defined as
\begin{equation}
  \mathcal{C}^{^{\perp_{_\mathrm{Tr}}}}:=\biggl\{\textbf{u}(\overline{x})\in\mathcal{S}_n \mid (\forall \textbf{c}(\overline{x})\in \mathcal{C})(\textbf{u}(\overline{x})\circledast \textbf{c}(\overline{x})=0) \biggr\}.  
\end{equation}
\end{defn}

Since $\circledast$-inner product is non-degenerate,   by Lemma \ref{lm-4.6}, we have the following result.

\begin{prop} Let $\mathcal{C}$ be a free $S|R$ additive code of length $n$. Then  $\mathcal{C}^{^{\perp_{_\mathrm{Tr}}}}$ is free (as $R$-module), $$\left(\mathcal{C}^{^{\perp_{_\mathrm{Tr}}}}\right)^{^{\perp_{_\mathrm{Tr}}}}=\mathcal{C},$$ and $\rk_R(\mathcal{C})+\rk_R(\mathcal{C}^{^{\perp_{_\mathrm{Tr}}}})=2n.$
\end{prop}

Let $\mathcal{C}$ be an $S|R$ additive code of length $n$. If $\textbf{u}(\overline{x})\in \mathcal{C}^{^{\perp_{_\mathrm{Tr}}}}$, $\textbf{c}(\overline{x})\circledast\textbf{u}(\overline{x}) = 0$ for all $\textbf{c}(\overline{x}) \in \mathcal{C}$. Since $\textbf{c}(\overline{x}) \in \mathcal{C}$, we know that $\overline{x}^{^{n-1}}\textbf{c}(\overline{x})$ is also a codeword. Thus, $$0=\overline{x}^{^{n-1}}\textbf{c}(\overline{x})\circledast\textbf{u}(\overline{x}) = \textbf{c}(\overline{x})\circledast\overline{x}\textbf{u}(\overline{x})$$ for all $\textbf{c}(\overline{x})$ from $\mathcal{C}$. Therefore $\overline{x}\textbf{u}(\overline{x}) \in \mathcal{C}^{^{\perp_{_\mathrm{Tr}}}}$ and $\mathcal{C}^{^{\perp_{_\mathrm{Tr}}}}$ is also an $S|R$ additive code of length $n$. Henceforth, we obtain the following proposition.

\begin{prop} Let $\mathcal{C}$ be an $S|R$ additive cyclic code of length $n$. Then $\mathcal{C}^{^{\perp_{_\mathrm{Tr}}}}$ is also an $S|R$ additive cyclic code of length $n$. 
\end{prop}

The $\circledast$-inner product is an $R$-bilinear form. On the other hand, $\mathcal{C}$ and $\mathcal{C}^{^{\perp_{_\mathrm{Tr}}}}$ are $\mathcal{R}_n$-submodules of the $\mathcal{R}_n$-module. Thus, we have the following result.

\begin{lemma}\label{rem-dual} Let $f(x), f'(x), g(x)$ and $g'(x)$ be monic divisors of $x^n-1$ over $R$ and $ (r(x), r'(x)) \in (R[x])^2 $ with $\deg(f(x)r(x))<\deg(g(x))$ and $\deg(f'(x)r'(x))<\deg(g'(x))$ such that 
$
\mathcal{C} = \langle f(\overline{x})(1 + \mu r(\overline{x})), \mu g(\overline{x}) \rangle_{_{\mathcal{R}_n}}$. Then
$$
\mathcal{C}^{^{\perp_{_\mathrm{Tr}}}} = \langle f'(\overline{x})(1 + \mu r'(\overline{x})), \mu g'(\overline{x}) \rangle_{_{\mathcal{R}_n}}
$$ if and only if  for all $0 \leq i,j < n$,
\begin{align}
    \overline{x}^if(\overline{x})\star  \overline{x}^jf'(\overline{x})=&-\mu^2(\overline{x}^if(\overline{x})r(\overline{x})\star \overline{x}^jf'(\overline{x})r'(\overline{x}));\tag{E1} \label{eq1}\\
   \overline{x}^ig(\overline{x})\star \overline{x}^jf'(\overline{x})r'(\overline{x})=&0;\tag{E2} \label{eq2} \\
    \overline{x}^if(\overline{x})r(\overline{x})\star \overline{x}^jg'(\overline{x})=&0;\tag{E3} \label{eq3}\\
   \overline{x}^ig(\overline{x}) \star \overline{x}^jg'(\overline{x})=&0\tag{E4} \label{eq4} .
\end{align}
Moreover $\deg(f(x))+\deg(f'(x))+\deg(g(x))+\deg(g'(x))=2n$.
\end{lemma}

\begin{example} Let $ R = \mathbb{Z}_9 $ and define $ S = R[\alpha] $, where $ \alpha^2 = -1 $. Consider the additive cyclic code
$
\mathcal{C} := \langle 1 + \alpha \overline{x} \rangle_{\mathcal{R}_n},
$
where $ n $ is a positive integer coprime to 3. 
The trace dual of $ \mathcal{C} $ is given by
\[
\mathcal{C}^{\perp_{\operatorname{Tr}}} = \left\langle f'(\overline{x})\big(1 + \alpha r'(\overline{x})\big),\ \mu g'(\overline{x}) \right\rangle_{\mathcal{R}_n},
\]
where $ f'(x)$ and $ g'(x)$ are monic divisors of $ x^n - 1 $ over $ R $, and $ r'(x) \in R[x] $ with $ \deg(r'(x)) < \deg(g'(x)) - \deg(f'(x)) $(since $ \operatorname{Tr}(\alpha) = 0 $).
We have $ \operatorname{rk}_R(\mathcal{C}) = \operatorname{rk}_R(\mathcal{C}^{\perp_{\operatorname{Tr}}}) = n = \deg(f'(x)g'(x)) $, and the following condition holds: For all $0 \leq i,j < n$
\begin{align}\label{LS0}
\overline{x}^i \star \overline{x}^{j} f'(\overline{x}) = \overline{x}^{i+1} \star \overline{x}^{j} f'(\overline{x}) r'(\overline{x}),
\text{ and } 
\overline{x}^i \star \overline{x}^j g'(\overline{x}) = 0.
\end{align}
From this, it follows that $ g'(x) = x^n - 1 $ and $ f'(x)= 1 $, so that $ \deg(r'(x)) < n $. Then, equation~\eqref{LS0} becomes: for all $0 \leq i,j < n$,
$
 \delta_{i,j} = \overline{x}^i \star \overline{x}^j = \overline{x}^{i+1} \star \overline{x}^{j} r'(\overline{x}).
$
This implies $ r'(\overline{x}) = \overline{x} $, and therefore $ \mathcal{C} $ is a trace self-dual additive cyclic code.
\end{example}

To determine the trace dual of an additive cyclic $R|S$ code of length $n$, we will define the following  polynomial operator:
\begin{equation}
    \begin{array}{crcl}
  ^\ast : & S[x]\backslash\{0\} & \rightarrow & S[x]\backslash\{0\} \\
     & {a}(x) & \mapsto & {a}^\ast(x)=x^{\deg({a}(x))}{a}(x^{-1}).
 \end{array}
\end{equation}
Note that if $ {a}(0)\neq 0$ then $\deg({a}(x))=\deg({a}^\ast(x)).$

\begin{lemma}\label{xn1}
   Let $  {a}(x)$ and ${b}(x)$ be two polynomials over $R$ of degree at most $n-1$. Then $x^n-1$ divides ${a}(x){b}(x)$ if and only if $${u}(\overline{x}){a}(\overline{x})\star{v}(\overline{x}){b}^*(\overline{x})=0$$ for any $( {u}(x), {v}(x))\in (R[x])^2$.
\end{lemma}

\begin{proof}
Let us prove $x^n-1$ divides ${a}(x){b}(x)$ if and only if ${u}(\overline{x}){a}(\overline{x})\star{v}(\overline{x}){b}^*(\overline{x})=0$, for any $( {u}(x), {v}(x))\in (R[x])^2$.

\emph{$\Rightarrow)$} Suppose $x^n - 1$ divides $ {a}(x) {b}(x)$. Then in the quotient ring $\mathcal{R}_n$, we have
$
\mathbf{a}(\overline{x})\mathbf{b}(\overline{x})= 0.
$
Multiplying on the left by any $ {u}(\overline{x})$ and on the right by any $ {v}(\overline{x})$, we get
$
 {u}(\overline{x}) {a}(\overline{x}) {b}(\overline{x}) {v}(\overline{x})=0.
$
That is,
\[
 {u}(x) {a}(x) {b}(x) {v}(x) = (x^n - 1) {q}(x),
\]
for some $\mathbf{q}(x) \in R[x]$.
Now consider
\[
({u}(\overline{x}){a}(\overline{x})) \star ({v}(\overline{x}){b}^*(\overline{x})) = \text{[constant term of]} \ {u}(x){a}(x)\cdot ({v}(x){b}^*(x^{-1})).
\]
But since ${b}^*(x^{-1}) = x^{-\deg {b}} {b}(x)$, it follows that
\[
{u}(x){a}(x)\cdot {v}(x){b}^*(x^{-1}) = x^{-\deg({b})} {u}(x){a}(x){b}(x){v}(x) = x^{-\deg {b}} (x^n - 1){q}(x).
\]
Since $(x^n - 1){q}(x)$ has no degree zero term, neither does $x^{-\deg {b}}(x^n - 1){q}(x)$. Thus, the constant coefficient is zero:
$
({u}(\overline{x}){a}(\overline{x})) \star ({v}(\overline{x}){b}^*(\overline{x})) = 0.
$

\emph{$\Leftarrow)$} Conversely, assume that for all ${u}(x), {v}(x) \in R[x]$,
\[
({u}(\overline{x}){a}(\overline{x})) \star ({v}(\overline{x}){b}^*(\overline{x})) = 0.
\]
We argue by contradiction. Suppose that ${a}(\overline{x}){b}(\overline{x}) =0$. Then ${a}(x){b}(x)$ is nonzero in $\mathcal{R}_n$, which is a free $R$-module of rank $n$.
The bilinear form
\[
(f,g) \mapsto (f(\overline{x}) \star g(\overline{x}))
\]
is non-degenerate on the free $R$-module $\mathcal{R}_n$. Hence, there exists some ${w}(x)$ such that
$
({w}(\overline{x}) \star {a}(\overline{x}){b}(\overline{x})) \neq 0.
$
Define ${u}(x) = 1$ and ${v}(x) = {w}(x){a}(x)x^{-\deg {b}}$, so that
$
{v}(x){b}^*(x^{-1}) = {w}(x){a}(x){b}(x),
$
and thus
\[
({u}(\overline{x}){a}(\overline{x})) \star ({v}(\overline{x}){b}^*(\overline{x})) = ({a}(\overline{x})) \star ({w}(\overline{x}){a}(\overline{x}){b}(\overline{x})) \neq 0.
\]
This contradicts the hypothesis, so it must be that ${a}(\overline{x}){b}(\overline{x}) = 0$, i.e., $x^n - 1 $ divides $ {a}(x){b}(x)$.
\end{proof}

The following result characterizes the trace dual of an additive cyclic code of length $n$ over $S|R$.

\begin{thm}\label{dual} Let $f(x), f'(x), g(x)$ and $g'(x)$ be monic divisors of $x^n-1$ over $R$ such that $$x^n-1=f(x)g_1(x)\ell(x)=f_1(x)g(x)\ell(x)$$  and $ (r(x), r'(x)) \in (R[x])^2 $ with $\deg(f(x)r(x))<\deg(g(x))$ and $\deg(f'(x)r'(x))<\deg(g'(x))$ such that 
$
\mathcal{C} = \langle f(\overline{x})(1 + \mu r(\overline{x})), \mu g(\overline{x}) \rangle_{_{\mathcal{R}_n}}$. Then
$$
\mathcal{C}^{^{\perp_{_\mathrm{Tr}}}} = \langle f'(\overline{x})(1 + \mu r'(\overline{x})), \mu g'(\overline{x}) \rangle_{_{\mathcal{R}_n}}
$$ if and only if 
\begin{align}
    f_1^*(x) \ell^*(x) \mid f'(x)r'(x) \text{ and } g_1'(x)\ell'(x)\mid  g^*(x)r'(x);\tag{R1} \label{r1}\\
    g_1^*(x) \ell^*(x) \mid  g'r^*(x)\text{ and } f_1'(x)\ell'(x)\mid  f^*(x)r^*(x);\tag{R2} \label{r2}\\
    f_1^*(x)\ell^*(x) \mid g'(x) \text{ and } f_1'(x)\ell'(x)\mid g^*(x)\tag{R3} \label{r3},
\end{align}
and for all $0 \leq i < n$,
$$
    \overline{x}^if(\overline{x})\star  f'(\overline{x})=-\mu^2(\overline{x}^if(\overline{x})r(\overline{x})\star f'(\overline{x})r'(\overline{x})).
$$ Moreover $ \ell^*(x)| f'(x)$. 
\end{thm}

\begin{proof} 
According to Remark~\ref{rem-dual} and Lemma~\ref{xn1}, we have for all $0 \leq i < n$,
$$
    \overline{x}^if(\overline{x})\star  f'(\overline{x})=-\mu^2(\overline{x}^if(\overline{x})r(\overline{x})\star f'(\overline{x})r'(\overline{x})).
$$   and the following equivalences:
\begin{align*}
    \eqref{eq2} \Leftrightarrow &\, x^n-1 \mid g^*(x) f'(x) r'(x) \Leftrightarrow f_1^*(x) \ell^*(x) \mid f'(x)r'(x)   \\ &\text{and } g_1'(x)\ell'(x)\mid  g^*(x)r'(x),  \eqref{r1} \\
    \eqref{eq3} \Leftrightarrow &\, x^n-1 \mid g'(x) f^*(x) r^*(x)\\  \Leftrightarrow &\, g_1^*(x) \ell^*(x) \mid  g'(x)r^*(x) \text{ and } f_1'(x)\ell'(x)\mid  f^*(x)r^*(x), \eqref{r2} \\
    \eqref{eq4} \Leftrightarrow & \, x^n-1 \mid g^*(x) g'(x) \\\Leftrightarrow &\, f_1^*(x) \ell^*(x) \mid g'(x) \text{ and } f_1'(x)\ell'(x)\mid g^*(x). \eqref{r3}
\end{align*} 
Finally, from Relation~\eqref{r1} and Equation~\eqref{eq1} one has that
\[ x^n-1 \mid f^*(x) g_1^*(x) f'(x) \Leftrightarrow \ell^*(x) \mid f'(x). \]
\end{proof}

\begin{example}
    Let $n = 8$, $p = 3$ and $S =\mathbb{Z}_9[\alpha]$ with $\alpha^2 =-1$ and $\mu=\alpha$. The factorization of $x^8-1$ into monic irreducible polynomial is given by 
    $$x^8-1=
    (x + 2) (x + 1) (x^2 + 1)(x^2 + x + 2)(x^2 + 2x + 2) \text{ in }  \mathbb{F}_3[x], $$
    and  the factorization of $x^8-1$ into monic basic irreducible polynomial over $\mathbb{Z}_9$ is given by: $$x^8-1=(x+8)(x+1)(x^2+1)(x^2+4x+8)(x^2+5x+8),$$
    
    Suppose $w(x) =x + 8$, $f_1(x)=x+1, $$g_1(x) =(x^2 + 4x + 8) (x^2 + 1)$, $\ell(x) = (x^2 + 5x + 8)$, and $r(x)=x^2+x+3$. Let \( \CC \) be an $\mathbb{Z}_9[\alpha]|\mathbb Z_9$ additive cyclic code of length 8 defined by  
\[
\begin{aligned}
\CC &= \big\langle w(\overline{x}) f_1(\overline{x})(1 + \mu  r(\overline{x})),\ \mu w(\overline{x}) g_1(\overline{x}) \big\rangle \\
  &= \big\langle (\overline{x}+8)(\overline{x}+1)(1+ 2\alpha(\overline{x}^2+\overline{x}+3)),\ 2\alpha(\overline{x} + 8)(\overline{x}^2 + 4\overline{x} +8) (\overline{x}^2 + 1)  \big\rangle \\
  &= \big\langle (\overline{x}^2+8)(1+2\alpha(\overline{x}^2+\overline{x}+3)),\ 2\alpha(\overline{x}^5+3\overline{x}^4+5\overline{x}^3+4\overline{x}^2+4\overline{x}+1 ) \big\rangle.
\end{aligned}
\]
The trace dual of  $\CC$ is given by  $\mathcal{C}^{^{\perp_{_\mathrm{Tr}}}} = \langle f'(\overline{x})(1 + \mu r'(\overline{x})), \mu g'(\overline{x}) \rangle_{_{\mathcal{R}_n}}$, where $x^8-1=w'(x)f_1'(x)g_1'(x)\ell'(x)=f'(x)g_1'(x)\ell'(x)=f_1'(x)g'(x)\ell'(x).$  In this case, Relations \eqref{r1} , \eqref{r2} and \eqref{r3} translate into 
\begin{itemize}
\item $(x+1)(x^2+4x+8) \mid  f'(x)r'(x)$ and  $g_1'(x)\ell'(x)\mid (x-1)(x^2 + 5x + 8) (x^2 + 1)r'(x)$;
\item 
$(x^2 + 5x + 8) (x^2 + 1)(x^2+4x+8)\mid (3x^2+x+1) g'(x)$ $  and  $  $f_1'(x)\ell'(x) \mid  (x^2-1)(3x^3+x+1)$;
\item
$(x+1)(x^2+4x+8) \mid    g'(x)$ and $f_1'(x)\ell'(x) \mid (x-1)(x^2 + 5x + 8) (x^2 + 1)$.
\end{itemize}
Thus
\begin{align*}
 (x+1)(x^2+4x+8)\mid &\;f'(x)r'(x) ; \\
 f_1'(x)\ell'(x)\mid &\; (x-1)(x+1) ; \\
  g_1'(x)\ell'(x)\mid &\;(x-1)(x^2 + 5x + 8) (x^2 + 1)r'(x);\\
\left(\frac{x^8-1}{w(x)}\right)\mid  &\;  g'(x).
\end{align*}
In addition $ (x^2+4x+8)\mid f'(x)$. Therefore $\deg(f'(x))\geq2$ and $g'(x)=\frac{x^8-1}{w(x)}$.  But $\deg(f'(x)g'(x))=9$ and $0\leq \deg(r'(x))<\deg(g'(x))-\deg(f'(x))$. Thus $f'(x)=x^2+4x+8, \ell'(x)=x-1$ and $0\leq \deg(r'(x))\leq 4$. Therefore $r'(x)=(x+1)(a_3x^3+a_2x^2+a_1x+a_0)$, where $(a_0,a_1, a_2, a_3)\in(\mathbb{Z}_9)^4$ and
\begin{align*} \overline{x}^i(\overline{x}^2-1)\star  (\overline{x}^2+4\overline{x}+8) = & \overline{x}^i(\overline{x}^4+\overline{x}^3+2\overline{x}^2+8\overline{x}+6) \\ &\star (\overline{x}^3+5\overline{x}^2 +3\overline{x}+8)(a_3\overline{x}^3+a_2\overline{x}^2+a_1\overline{x}+a_0),
\end{align*}
for all $0\leq i\leq 7.$
Thus, we get $(a_0,a_1, a_2, a_3)=(6,4,3,5)$.
\end{example}
The following result states that for the trace dual code to be stated in terms of the reciprocal polynomials the remainder $r(x)$ should be 0.
\begin{cor}\label{cor-4.5} Let $f(x), f'(x), g(x)$ and $g'(x)$ be monic divisors of $x^n-1$ in $R[x]$ and $ (r(x), r'(x)) \in (R[x])^2 $ with $\deg(f(x)r(x))<\deg(g(x))$ and $\deg(f'(x)r'(x))<\deg(g'(x))$ such that 
$
\mathcal{C} = \langle f(\overline{x})(1 + \mu r(\overline{x})), \mu g(\overline{x}) \rangle_{_{\mathcal{R}_n}}$ and
$$
\mathcal{C}^{^{\perp_{_\mathrm{Tr}}}} = \langle f'(\overline{x})(1 + \mu r'(\overline{x})), \mu g'(\overline{x}) \rangle_{_{\mathcal{R}_n}}
$$
 Then  $f'(x)=\left(\frac{x^n-1}{f(x)}\right)^*$ and $g'(x)=\left(\frac{x^n-1}{g(x)}\right)^*$, if and only if $r(x)=r'(x)=0$.
\end{cor}

\begin{proof} Let $f(x)=w(x)f_1(x)$ and $g(x)=w(x)g_1(x)$ where $f_1(x)$ and $g_1(x)$ are coprime with $x^n-1=f(x)g_1(x)\ell(x)=f_1(x)g(x)\ell(x)$. Assume that $f'(x)=\left(\frac{x^n-1}{f(x)}\right)^*=g_1^*(x)\ell^*(x)$ and $g'(x)=\left(\frac{x^n-1}{g(x)}\right)^*=f_1^*(x)\ell^*(x)$. By Theorem~\ref{dual},  
we have $f_1^*(x)\ell^*(x)|f^*(x)r^*(x)$ and $ w^*(x)g_1^*(x)|g_1^*(x)\ell^*(x)r'(x)$.  Thus $f_1^*(x)$ divides $r'(x)$ and $g_1(x)$ divides $r(x)$. Hence $r(x)=r'(x)=0$, since $\deg(r(x))< \deg(g_1(x))-\deg(f_1(x))$ and $\deg(r'(x))< \deg(g'_1(x))-\deg(f'_1(x))$. The converse is a direct consequence of Lemma~\ref{xn1}, and the fact that $\deg(f(x))+\deg(f'(x))+\deg(g(x))+\deg(g'(x))=2n$.
\end{proof}

\section{Additive complementary pairs of codes}\label{sec:acp}

In this section, we focus on Additive Complementary Pairs (ACP) over finite commutative chain rings.

\begin{defn}\label{def:acp}
Let $\CC$ and $\CD$ be two $ S | R $ additive codes. If $\CC+_{R}\CD=S^n$ and $\CC\cap\CD=\{\mathbf 0\}$, then the pair $\{\CC, \CD\}$ is called an ACP of codes.     
\end{defn}

\begin{remark}\label{rm-2}
 As usual, we will denote the conditions in Definition~\ref{def:acp} as $\CC\oplus_{R} \CD=S^n$.   Note that for cyclic codes, we can use the identification of $S^n$ with $\mathcal{S}_n$ and we can say that (taking into account now that the codes are ideals in $\mathcal{S}_n$) then they are an ACP if $\CC\oplus_{R} \CD=\mathcal{S}_n$.
\end{remark}


\begin{lemma}\label{kap}\cite[Theorem 2]{Kap58}
Any projective module over a local ring is free module.
\end{lemma}
\begin{lemma}\label{lem:free} Let $\CC$ and $\CD$ be two additive codes over $ S | R $. If the pair $\{\CC,\CD\}$ is an ACP of codes, then both  $\CC$ and $\CD$ are free $R$-modules of $S^n$.
\end{lemma}
\begin{proof}
 Since the pair $\{\CC,\CD\}$ forms an ACP of codes, then by Remark~\ref{rm-2}, we get $\CC\oplus_{R} \CD=S^n$ it follows that $C\oplus D$ is free  $R$-module. This implies that $\CC$ and $\CD$ both are projective $R$-submodule of $S^n$. Since $R$ and $S$ are local rings, then by Lemma~\ref{kap}, $\CC$ and $\CD$ both are free $R$-submodule of $S^n$.
\end{proof}

Now, the result \cite[Lemma 3.1]{BD24} easily adapts to the case of additive codes.

\begin{lemma} Let $\CC$ and $\CD$ be two $ S | R $ additive codes. Then we have the following theorem.
\begin{enumerate}
\item $(\CC+\CD)^{\perp_{\mathrm{Tr}}}=\CC^{\perp_{\mathrm{Tr}}}\cap \CD^{\perp_{\mathrm{Tr}}}$;
\item $\CC^{\perp_{\mathrm{Tr}}}+\CD^{\perp_{\mathrm{Tr}}}=(\CC\cap \CD)^{\perp_{\mathrm{Tr}}}$.
 \end{enumerate}
\end{lemma}

\begin{proof} $\quad$
\begin{enumerate}
\item Let $\mathbf x\in (\CC+\CD)^{\perp_{\mathrm{Tr}}}$. Then
$\mathrm{Tr}( \mathbf x, \mathbf a)=0$  for all $\mathbf a\in \CC+\CD$, that is   $\mathrm{Tr}( \mathbf x, \mathbf c+ \mathbf d)=0$ for all $\mathbf c\in \CC, \mathbf d\in \CD.$
If $ \mathbf d = 0$ then $\mathrm{Tr}(\mathbf x, \mathbf c)=0$ for all $\mathbf c\in \CC$, then $\mathbf x\in \CC^{\perp_{\mathrm{Tr}}}$. Similarly, if $ \mathbf c = 0$ then $\mathrm{Tr}(\mathbf x, \mathbf d)=0$ for all $\mathbf d\in \CD$, which implies  $\mathbf x\in \CD^{\perp_{\mathrm{Tr}}}$. Hence, $\mathbf x\in \CC^{\perp_{\mathrm{Tr}}}\cap \CD^{\perp_{\mathrm{Tr}}}.$

On the other hand, let $ \mathbf y \in  \CC^{\perp_{\mathrm{Tr}}}\cap \CD^{\perp_{\mathrm{Tr}}}$. 
Then $\mathrm{Tr}(\mathbf y,\mathbf c)=0$ for~all $\mathbf c\in \CC$ and $\mathrm{Tr}( \mathbf y, \mathbf d)=0$ for~all $\mathbf d\in \CD$.  That implies $\mathrm{Tr}(\mathbf y, \mathbf c+\mathbf d)=\mathrm{Tr}(\mathbf y, \mathbf c)+\mathrm{Tr}(\mathbf y,\mathbf d)=0$ for~all $ \mathbf c\in \CC,~ \mathbf d\in \CD$. Hence, $\mathbf y\in (\CC+\CD)^{\perp_{\mathrm{Tr}}}$.
\item As $\CC^{\perp_{\mathrm{Tr}}}+\CD^{\perp_{\mathrm{Tr}}}=\left( (\CC^{\perp_{\mathrm{Tr}}}+\CD^{\perp_{\mathrm{Tr}}})^{\perp_{\mathrm{Tr}}} \right)^{\perp_{\mathrm{Tr}}}$, hence $$\CC^{\perp_{\mathrm{Tr}}}+\CD^{\perp_{\mathrm{Tr}}} = (\CC\cap \CD)^{\perp_{\mathrm{Tr}}}.$$
\end{enumerate}
\end{proof}
Taking into account the previous lemma, we have
\begin{thm}\label{th-1.1}
Let $\CC$ and $\CD$ be two additive codes over $ S | R $. Then the following statements  are equivalent.
\begin{enumerate}
\item the pair $(\CC,\CD)$ is an ACP of codes;
\item the pair $(\CC^{\perp_{\mathrm{Tr}}},\CD^{\perp_{\mathrm{Tr}}})$ is an ACP of codes.
\end{enumerate}
\end{thm}
The following result was shown for the linear codes in \cite{BTP24}. Here, we state it for an additive code and will allow us to characterize the ACP of additive $ S | R $-codes in terms of their ranks.
\begin{lemma}\label{lm-4.6}
 Let $\CC$ and $\CD$ be two free $ S | R $ additive codes. Denote $\CC+_{R}\CD=\langle\CC\cup\CD\rangle_{_{R}}$ that is the smallest $R$-submodule of $S^n$ containing $\CC\cup\CD$. Then \[
 \rk_{R}(\CC+_{R}\CD)=\rk_{R}(C)+\rk_{R}(\CD)-\rk_{R}(\CC\cap\CD).
 \] 
\end{lemma}
\begin{proof} To prove the result, consider the map
 $$
\begin{array}{cccc}
  \varphi : & \CC\times \CD & \mapsto & \CC+_{\mathcal{\mathcal{R}}}\CD \\
    & (x, y) & \mapsto & x+ y.
\end{array}
 $$
 Obviously this map $\varphi$ is an $R$-module homomorphism. Then by Remark~\ref{rm-2}, $\CC+_{R}\CD$ is also $R$-module. It is clear that the map $\varphi$ is surjective. Therefore, according to the First Isomorphism Theorem, $\CC\times \CD/\mathrm{Ker}(\phi)\simeq  \CC+_{R}\CD$ (as $R$-modules). Since $\CC\cap \CD \simeq \mathrm{Ker}(\phi)$ (as $R$-modules), it follows that $\frac{|\CC\times \CD|}{|\CC\cap \CD|}=|\CC+_{R}\CD|$. Thus, $\rk_{R}(\CC+_{R}\CD)=\rk_{R}(C)+\rk_{R}(\CD)-\rk_{R}(\CC\cap\CD).$
 \end{proof}
 
\begin{cor}\label{th-4.7}
Let $\CC$ and $\CD$ be two $S | R $ additive codes of length $n$. Then $\{\CC, \CD\}$ is an ACP of codes if and only if $\rk_{R}(\CC+_{R}\CD)=\rk_{R}(\CC)+\rk_{R}(\CD)=2n$.    
\end{cor}

\begin{proof}
By Lemma~\ref{lem:free} $\CC,\CD$ are free $R$-modules since they form an ACP of codes. Thus, applying the lemma~\ref{lm-4.6}, we easily deduce that $\{\CC, \CD\}$ is an ACP of codes if and only if $\rk_{R}(\CC+_{R}\CD)=\rk_{R}(\CC)+\rk_{R}(\CD)=2n$.
\end{proof}

Let $\pi$ the natural surjective ring homomorphism  
$\pi : S \rightarrow S/\mathfrak{m}_{S}=\mathbb{F}_{q^2}$  which naturally extends to a homomorphism from $ S^n $ to $\mathbb{F}_{q^2}^n$.

\begin{lemma}\label{p-4.1}(\cite[Nakayama's Lemma]{McDonal1974})
Let \(M\) be a finitely generated $R$-module. If  
$\mathfrak{m}_{R}M = M,$
then \(M = 0\).
\end{lemma}

\begin{lemma}\label{lm-4.9}
  Let $\CC$ and $\CD$ be two $S | R $ additive  codes. Then $\CC \cap \CD = \{\mathbf 0\}$ if and only if $\pi(C) \cap \pi(D) = \{\mathbf 0\}$.  
\end{lemma}
\begin{proof}
Assume that $ \pi(\CC) \cap \pi(\CD) = \{\mathbf 0\} $.  
Let $\mathbf v \in \CC \cap \CD $. Then $\pi(\mathbf v) \in \pi(\CC) \cap \pi(\CD) $ and, by hypothesis, $ \pi(\mathbf v) = 0 $. Therefore, $\mathbf v \in \mathfrak{m}_{R}(\CC \cap \CD) $, which implies
$ 
\CC \cap \CD = \mathfrak{m}_{R}(\CC \cap \CD).
$ 
Since $ \CC \cap \CD $ is a finitely generated right \( R \)-module, Proposition \ref{p-4.1} implies that \( \CC \cap \CD = \{\mathbf 0\} \).

Conversely, assume that $ \CC \cap \CD = \{\mathbf 0\} $ and let $\mathbf v \in \pi(\CC) \cap \pi(\CD) $. Then there exist $ \mathbf c \in \CC $ and $ \mathbf d \in \CD $ such that $ \pi(\mathbf c) = \pi(\mathbf d) = \mathbf v $. Hence, $ \pi(\mathbf c - \mathbf d) = 0 $, which implies that $\mathbf c - \mathbf d \in \mathfrak{m}_{R}S^n$. Hence, there is a power $\gamma^i$ of $\gamma$  (the generator of $\mathfrak{m}_{R}$)  with $1\leq i<e$ such that   $\gamma^i(\mathbf c -\mathbf d)= \mathbf 0 $, then  $ \gamma^i \mathbf c = \gamma^i \mathbf d  \in C \cap D $. By assumption, $ \CC \cap \CD = \{\mathbf 0\} $, so $  \gamma^i \mathbf c = \gamma^i \mathbf d =\mathbf 0 $.  
Suppose $\mathbf c \notin \mathfrak{m}_{R}S^n$. Then $ \gamma^i \mathbf c  \neq \mathbf 0 $ which contradicts the previous statement. Thus, $ \mathbf c \in \mathfrak{m}_{R}S^n$, implying $\pi(\mathbf c) = \mathbf 0$, and hence $ \mathbf v = \pi(\mathbf c) = \mathbf 0 $. This shows that $ \pi(\CC) \cap \pi(\CD) = \{\mathbf 0\} $.   
\end{proof}
\begin{thm}\label{th-4.10}
 Let $\CC$ and $\CD$ be two $ S | R $  additive codes of length $n$. The pair $ \{\CC, \CD\} $ forms an ACP of codes if and only if the pair $\{\pi(\CC), \pi(\CD)\}$ also forms an ACP of $\mathbb F_{q^2}|\mathbb F_q$-linear codes.    
\end{thm}

\begin{proof} $\quad$
\begin{itemize}
    \item[] $\Rightarrow )$ Since $ \{\CC, \CD\} $ is an ACP, we have $ \CC \cap \CD = \{\mathbf 0\} $, thus by Lemma~\ref{lm-4.9}, it follows that $\pi(C) \cap \pi(D) = \{\mathbf 0\} $.    
Let $ \mathbf v \in \mathbb{F}_{q^2}^n $, Since $ \pi $ is a subjective map, there exists $ \mathbf a \in S^n $ such that $\pi(\mathbf a) = \mathbf v$.  
Since $\CC +_{R} \CD = S^n $, there are $ \mathbf c \in \CC $, $\mathbf d \in \CD $ such that $\mathbf a = \mathbf c + \mathbf d $.  
Henceforth, $ \mathbf v = \pi(\mathbf a) = \pi(\mathbf c) + \pi(\mathbf d) \in \pi(\CC) + \pi(\CD) $.  
Therefore, $ \pi(\CC) +_{\FF_q} \pi(\CD) = \mathbb{F}_{q^2}^n $, and we conclude that $(\pi(C), \pi(D)) $ is ACP.   
\item[] $\Leftarrow )$ Suppose now that $(\pi(C), \pi(D)) $ is ACP  of $\mathbb F_{q^2}|\mathbb F_q$-linear codes. Then
$ 
\pi(\CC) \oplus_{\FF_q} \pi(\CD) = \mathbb{F}_q^n$, which implies   $\pi(\CC) \cap \pi(\CD) = \{\mathbf 0\}$ . Thus,
by Lemma~\ref{lm-4.9}, it follows that $\CC \cap \CD = \{\mathbf 0\} $.

Let $ \{\pi(\mathbf x_1), \ldots, \pi(\mathbf x_k)\} $ be a basis of $\pi(\CC) $, and $\{\pi(\mathbf x_{k+1}), \dots, \pi(\mathbf x_n)\}$  a basis of $ \pi(\CD) $. Then, it is straightforward that $ \{\mathbf x_1, \dots, \mathbf x_k\} $ and $\{\mathbf x_{k+1}, \dots,\mathbf x_n\}$ are minimal generating sets for $ \CC $ and $ \CD $, respectively.
Since $ \CC $ and $ \CD $ are both free, we have
$|\CC||\CD| = |S^n|.$
Therefore, $ \CC +_{R} \CD = S^n $, and the pair $\{\CC, \CD\} $ is an ACP of additive codes of length $n$.
\end{itemize}
\end{proof}

Let $ n $ be a positive integer. An $ n \times n $ matrix $ \mathrm{A} $ over $ \mathcal{S} $ is said to be invertible over $ \mathcal{S} $ if the matrix $ \pi(\mathrm{A}) = (\pi(a_{ij})) $ is invertible over $\mathbb{F}_{q^2} $.  For a free code $C$,
 consider a generator matrix $\mathrm{G}$ and parity-check matrix $\mathrm{H} $ of $\CC$. Then, under the projection $ \pi $, the matrices $\pi(\mathrm{G}) $ and  $\pi(\mathrm{H}) $ serve as a generator matrix and a parity-check matrix of the projected code $ \pi(\CC) $, respectively.

\begin{prop}\cite[Theorem 3.7]{BD24}\label{p-4.4}
 Let $\CC$ and $\CD$ be two free $ S | R $ additive codes of length $n$   with generator matrices \( \mathrm{G}_1 \), \( \mathrm{G}_2 \) and parity-check matrices \( \mathrm{H}_1 \), \( \mathrm{H}_2 \), respectively. Suppose that   $|\CC|  |\CD| = |S^n|$.  
 Then, the following statements are equivalent:
 \begin{enumerate}
     \item[a)] The pair \( \{\pi(\CC), \pi(\CD)\} \) forms ACP,
     \item[b)] The matrix $\mathrm{Tr}\left(\pi(\mathrm{H}_2)\pi(\mathrm{G}_1)^\top\right)$ or $\mathrm{Tr}\left(\pi(\mathrm{H}_1)\pi(\mathrm{G}_2)^\top\right) $ is invertible over $\mathbb{F}_{q^2} $.
 \end{enumerate}
\end{prop}
\begin{thm}
 Let $\CC$ and $\CD$ be two free additive codes of length $n$ over $ S | R $ with generator matrices $ \mathrm{G}_1 $, $ \mathrm{G}_2 $ and parity-check matrices  $\mathrm{H}_1 $, $\mathrm{H}_2 $, respectively. Then the pair $\{\CC, \CD\} $ is ACP if and only if $\mathrm{Tr}\left(\mathrm{H}_2 \mathrm{G}_1^\top \right)$ or $\mathrm{Tr}\left(\mathrm{H}_1 \mathrm{G}_2^\top \right)$ is invertible.   
\end{thm}
\begin{proof}
 Suppose $\{\CC, \CD\}$  is ACP of  free additive codes over $ S | R $. Assume, for contradiction, that $\mathrm{Tr}\left(\mathrm{H}_2 \mathrm{G}_1^\top \right)$ is not invertible. Then its image under the ring homomorphism $ \pi $ $$\pi (\mathrm{Tr}\left(\mathrm{H}_2 \mathrm{G}_1^\top\right)) = \mathrm{Tr}\left(\pi(\mathrm{H}_2)\pi(\mathrm{G}_1)^\top\right),$$ is also not invertible over $ \mathbb{F}_{q^2} $.   

 By Proposition \ref{p-4.4}, this implies $ \{\pi(\CC), \pi(\CD)\} $ is not ACP. Then, by applying  Theorem \ref{th-4.10}, $ (C, D) $ cannot be an ACP, contradicting our assumption. Hence, $\mathrm{Tr}\left(\mathrm{H}_2 \mathrm{G}_1^\top \right)$ or $\mathrm{Tr}\left(\mathrm{H}_1 \mathrm{G}_2^\top \right)$ must be invertible.

 Conversely, assume $\mathrm{Tr}\left(\mathrm{H}_2\mathrm{G}_1^\top \right)$ or $\mathrm{Tr}\left(\mathrm{H}_1 \mathrm{G}_2^\top \right)$ is invertible. Then the matrix $  \mathrm{Tr}\left(\pi(\mathrm{H}_2)\pi(\mathrm{G}_1)^\top\right)$ is invertible, implying $ \{\pi(\CC), \pi(\CD)\} $ is an ACP by Proposition \ref{p-4.4}. Therefore, by Theorem \ref{th-4.10}, $ \{\CC, \CD\} $ is ACP.
\end{proof}

\section{ACP of additive cyclic codes}\label{sec:acpc}

In this Section, we study pairs $\{\CC, \CD\}$  that are ACP of additive $ S | R $ cyclic codes. Note that by Lemma~\ref{lem:free}, both codes $\CC$ and $\CD$ are free additive codes over $S|R$ with a representation as in Theorem~\ref{th-3.3}.

\begin{thm}\label{main} Let $f_1(x), f_2(x), g_1(x)$ and $g_2(x)$ be monic divisors of $x^n-1$ over $R$ and $ (r_1(x), r_2(x)) \in (R[x])^2 $ with $\deg(f_1(x)r_1(x))<\deg(g_1(x))$ and $\deg(f_2(x)r_2(x))<\deg(g_2(x))$ such that 
$$
\mathcal{C} = \langle f_1(\overline{x})(1 + \mu r_1(\overline{x})), \mu g_1(\overline{x}) \rangle_{_{\mathcal{R}_n}}
\text{ and }
\mathcal{D}  = \langle f_2(\overline{x})(1 + \mu r_2(\overline{x})), \mu g_2(\overline{x}) \rangle_{_{\mathcal{R}_n}}
$$
 are two $ S | R $ additive cyclic codes of length $ n $. Then,   the pair $(\mathcal{C}, \mathcal{D})$ is ACP if and only if $f_1(x)f_2(x)=g_1(x)g_2(x)=x^n-1$.
\end{thm}

\begin{proof}
Assume that $\{\CC, \CD\}$ is ACP. By Theorem \ref{th-4.10} we know that  $\{\pi(\CC), \pi(\CD)\}$ is an ACP of $\mathbb F_q$-linear $\mathbb F_{q^2}$-codes.     Then $\langle \pi(f_1(x))\rangle_{_{\mathbb F_q}}+\langle \pi(f_2(x)) \rangle_{_{\mathbb F_q}}=\mathbb F_{q^2}^n$ and 
$$
 \langle \mu \pi(g_1)(\overline{x}) \rangle_{_{\mathbb F_q}}\cap\langle \mu \pi(g_2)(\overline{x}) \rangle_{_{\mathbb F_q}}\subseteq\pi(\CC)\cap\pi(\CD)=\{0\}.
$$
Thus $\mathrm{gcd}(\pi(f_1(x)), \pi(f_2(x)))=1,$ and $\lcm(\pi(g_1(x)),\pi(g_2(x))))=x^n-1$. It follows that $\pi(f_1(x))\pi(f_2(x))=\pi(g_1(x))\pi(g_2(x))=x^n-1$, since $f_1(x), f_2(x), g_1(x)$ and $g_2(x)$ are monic divisors of $x^n-1$ and $\deg(f_1(x)f_2(x))+\deg(g_1(x)g_2(x))=2n$. Hence, by Hensel lift, we have $f_1(x)f_2(x)=g_1(x)g_2(x)=x^n-1$.

Conversely, suppose that $f_1(x)f_2(x)=g_1(x)g_2(x)=x^n-1$. To prove the result, let us define a map
$$\begin{array}{cccc}
  \psi: & \CC\cap\CD & \rightarrow & \mathcal{R}_n \\
    & a(\overline{x})+\mu b(\overline{x}) & \mapsto & a(\overline{x}).
\end{array}$$   
 Clearly, $\CC\cap \CD$ is additive $R|S$ cyclic code of length $ n $. Obviously, $\texttt{Im}(\psi)=\langle f_1(x)f_2(x)\rangle$ and $\mathrm{Ker}(\psi)=\langle g_1(x)g_2(x)\rangle$. By Theorem~\ref{th-3.3}, we have
\[
\CC\cap\CD=\langle f_1(\overline{x})f_2(\overline{x})(1+\mu r(\overline{x})), \mu g_1(\overline{x})g_2(\overline{x}) \rangle_{_{\mathcal{R}_n}},
\] 
where $r(x)\in R[x]$. By hypothesis $f_1(\overline{x})f_2(\overline{x})=g_1(\overline{x})g_2(\overline{x})=\overline{x}^n-1=0$. 
Thus, we obtain $\CC\cap\CD=\{\mathbf 0\}$. Since $x^n-1$ is square-free, we get $\langle f_1(\overline{x}) \rangle+\langle f_2(\overline{x}) \rangle=\mathcal{R}_n$ ($f_1$ and $f_2$ are coprime) and $\langle g_1(\overline{x}) \rangle+\langle g_2(\overline{x})  \rangle=\mathcal{R}_n$ ($g_1(x)$ and $g_2(x)$ are coprime polynomials) which results $\rk_{_R}(\CC)+\rk_{_R}(\CD)=2n$. Therefore, applying Theorem~\ref{th-4.7}, the result follows.
\end{proof}

\begin{cor}\label{cor:norem}
      Let $f_1(x), f_2(x), g_1(x)$ and $g_2(x)$ are monic divisors of $x^n-1$ over $R$ and $ (r_1(x), r_2(x)) \in (R[x])^2 $ with $\deg(f_1(x)r_1(x))<\deg(g_1(x))$ and $\deg(f_2(x)r_2(x))<\deg(g_2(x))$ such that 
$$
\mathcal{C} = \langle f_1(\overline{x})(1 + \mu r_1(\overline{x})), \mu g_1(\overline{x}) \rangle_{_{\mathcal{R}_n}}
\text{ and }
\mathcal{D}  = \langle f_2(\overline{x})(1 + \mu r_2(\overline{x})), \mu g_2(\overline{x}) \rangle_{_{\mathcal{R}_n}}
.$$  If  $\{\mathcal{C},\mathcal{D}\}$ is ACP then $r_1(x)=r_2(x)=0$.
\end{cor}
\begin{proof}
    According to Theorem \ref{main},  it follows that $f_1f_2=g_1g_2=x^n-1$. Consider this map $$
 \begin{array}{cccc}
   \varphi : & \mathcal{S}_n & \rightarrow & \mathcal{R}_n \\
     & a(\overline{x})+\mu b(\overline{x}) & \mapsto & b(\overline{x}) 
 \end{array}
 $$ that is an epimorphism of $\mathcal{R}_n$-modules. We have $$\varphi(\mathcal{C})=\langle f_1(\overline{x}) r_1(\overline{x})),  g_1(\overline{x}) \rangle_{_{\mathcal{R}_n}}\text{ and  }\varphi(\mathcal{D})=\langle f_2(\overline{x}) r_2(\overline{x})),  g_2(\overline{x}) \rangle_{_{\mathcal{R}_n}}.$$ Note that $\mathcal{R}_n$ is a principal ideal ring, thus there exist polynomials $d_1(x)$ and $d_2(x)$  monic divisors of $x^n-1$ such that $\varphi(\mathcal{C})=\langle d_1(\overline{x})  \rangle_{_{\mathcal{R}_n}}$ and  $\varphi(\mathcal{D})=\langle d_2(\overline{x})  \rangle_{_{\mathcal{R}_n}}.$ Since $\mathcal{C}+\mathcal{D}=\mathcal{S}_n$ and $\varphi$ is an epimorphism of $\mathcal{R}_n$-modules, it follows that $\varphi(\mathcal{C})+\varphi(\mathcal{D})=\mathcal{R}_n$. Thus $d_1(x)$ and $d_2(x)$ are coprime polynomials and $d_1(x)d_2(x)$ divides $x^n-1$. According to Lemma \ref{lm-4.6}, we have $$\rk_{R}(\mathcal{R}_n)=\rk_{R}(\varphi(\mathcal{C}))+\rk_{R}(\varphi(\mathcal{D}))-\rk_{R}(\varphi(\mathcal{C})\cap\varphi(\mathcal{D})).$$ Thus $\deg(d_1(x))+\deg(d_2(x))=n$. Hence $d_1(x)d_2(x)=x^n-1$ and  $d_1(x)=g_1(x)$ and $d_2(x)=g_2(x)$.  Since $\deg(f_1(x)r_1(x))<\deg(g_1(x))$ and $\deg(f_2(x)r_2(x))<\deg(g_2(x))$, we have $f_1(x)r_2(x)=f_2(x)r_2(x)=0$.  It follows that $r_1(x)=r_2(x)=0$, since $f_1(x)$ and $f_2(x)$. 
\end{proof}

\begin{cor}\label{cor:per} Let $f_1(x), f_2(x), g_1(x)$ and $g_2(x)$ be monic divisors of $x^n-1$ over $R$  such that 
$\mathcal{C} = \mathcal{C}_1 \ \oplus \mu\mathcal{C}_2  
\text{ and }
\mathcal{D}  = \mathcal{D}_1 \ \oplus \mu\mathcal{D}_2
,$
where $\mathcal{C}_1=\langle f_1(\overline{x}) \rangle, \mathcal{C}_2=\langle g_1(\overline{x}) \rangle, \mathcal{D}_1=\langle f_2(\overline{x}) \rangle,$ and $\mathcal{D}_2=\langle g_2(\overline{x}) \rangle.$
 Let $\sigma$ be the permutation of $\{0,1,\cdots, n-1\}$ defined by $\sigma(i)=n-i-1$. Then the following assertions are equivalent.
\begin{enumerate}
  \item The pair $\{\mathcal{C},\mathcal{D}\}$ is ACP.
  \item The pairs $\{\mathcal{C}_1,\mathcal{D}_1\}$ and $\{\mathcal{C}_2,\mathcal{D}_2\}$ are LCP of codes.
  \item $\mathcal{C}_1^{^{\perp}}=\sigma(\mathcal{D}_1)$ and $\mathcal{C}_2^{^{\perp}}=\sigma(\mathcal{D}_2)$.
  \item $\mathcal{C}^{^{\perp_{\mathrm{Tr}}}}=\sigma(\mathcal{D})$. 
\end{enumerate}
\end{cor}

\begin{proof} $\quad$
\begin{description}
  \item[1) $\Rightarrow$ 2):] Assume $\{\mathcal C,\mathcal D\}$ is an ACP.  By Theorem~\ref{main}, we have
$
  f_1(x)f_2(x) \;=\; g_1(x) g_2(x) \;=\; x^n - 1.
$ Since $\gcd(f_1(x),f_2(x))=1$ (and respectively $\gcd(g_1(x),g_2(x))=1$), the cyclic codes
$
  \mathcal C_1=\langle f_1(\bar x)\rangle,
  $ and $\mathcal D_1=\langle f_2(\bar x)\rangle$
satisfy
\[
  \mathcal C_1\cap\mathcal D_1
    =\langle \operatorname{lcm}(f_1,f_2)\rangle
    = \{ 0\},
  \quad
  \mathcal C_1+\mathcal D_1
    =\langle \gcd(f_1,f_2)\rangle
    = R^n.
\]
Thus $\{\mathcal C_1,\mathcal D_1\}$ is an LCP.  Similarly $\{\mathcal C_2,\mathcal D_2\}$ is an LCP.
  \item[2) $\Rightarrow$ 3):] If $\{\mathcal C_i,\mathcal D_i\}$ is LCP, then $
  \mathcal C_i+\mathcal D_i=R^n,$ and $\mathcal C_i\cap\mathcal D_i=\{\textbf 0\},$
and for cyclic codes this is equivalent to
$f_i(x)\,g_i(x)=x^n-1.$ But by standard duality for cyclic codes, the Euclidean dual of $\langle f_i(\bar x)\rangle$ is
\[
  \langle f_i(\bar x)\rangle^\perp
    = \bigl\langle (x^n-1)/f_i^*(\bar x)\bigr\rangle,
\]
and coefficient-reversal via $\sigma(i)=n-1-i$ satisfies
$
  \langle f_i(\bar x)\rangle^\perp
    = \sigma\bigl(\langle g_i(\bar x)\rangle\bigr).
$
Hence $\mathcal C_i^\perp = \sigma(\mathcal D_i),$ for  $i=1,2.$
  \item[3) $\Rightarrow$ 4):] Assuming $\mathcal C_i^\perp=\sigma(\mathcal D_i)$ for $i=1,2$. Then  
\[
  \mathcal C^{\perp_{\mathrm{Tr}}}
    = \mathcal C_1^\perp \;\oplus\;\mu\,\mathcal C_2^\perp
    = \sigma(\mathcal D_1)\oplus\mu\,\sigma(\mathcal D_2)
    = \sigma(\mathcal D_1\oplus\mu\,\mathcal D_2)
    = \sigma(\mathcal D).
\]
  \item[4) $\Rightarrow$ 1):]  According to Corollary \ref{cor-4.5},  we have $$
\mathcal{C}^{^{\perp_{_\mathrm{Tr}}}} = \left\langle \left(\frac{x^n-1}{f_1(x)}\right)^*, \mu \left(\frac{x^n-1}{g_1(x)}\right)^* \right\rangle_{_{\mathcal{R}_n}}.
$$ On the other hand, $\sigma(\mathcal{D})=\left\langle f_2^*(x), \mu g_2^*(x) \right\rangle_{_{\mathcal{R}_n}}.$ Since $\mathcal{C}^{^{\perp_{_\mathrm{Tr}}}}=\sigma(\mathcal{D})$, by identification, it follows that $f_1(x)f_2(x)=g_1(x)g_2(x)=x^n-1$. Hence, by Theorem \ref{main}, the pair $\{\mathcal{C},\mathcal{D}\}$ is ACP. 
\end{description}

\end{proof}

Taking into account Collorary~\ref{cor:norem} and Corollary~\ref{cor:per} we have that the result of \cite{CG18} is also true for additive $S|R$ cyclic codes. 

\begin{thm} Let  the pair $\{\mathcal{C},\mathcal{D}\}$ be a ACP of $S|R$ cyclic codes of length $n$ and $\sigma$ be the permutation of $\{0,1,\cdots, n-1\}$ defined by $\sigma(i)=n-i-1$. Then $$\mathcal{C}^{^{\perp_{\mathrm{Tr}}}}=\sigma(\mathcal{D}).$$
\end{thm}

\bibliographystyle{plain} 
\bibliography{references} 

\end{document}